\documentclass[10pt,twocolumn,twoside] {IEEEtran}

\usepackage{cite}

\ifCLASSINFOpdf

\else
   \usepackage[dvips]{graphicx}
   \DeclareGraphicsExtensions{.eps}
\fi
\usepackage[cmex10]{amsmath}
\usepackage{array}
\usepackage{amsthm}
\theoremstyle{remark}

\usepackage{booktabs}
\usepackage{bm}
\usepackage{tabularx}

\usepackage{color} % pb_ll
\usepackage{booktabs}% pb_ll
\usepackage{bm}% pb_ll
\usepackage{tabularx}% pb_ll
\usepackage{amssymb} %pb_ll
\usepackage{algpseudocode}  %pb_ll

\begin{document}
\title{ Joint DOA and Frequency Estimation with Sub-Nyquist Sampling for More Sources than Sensors}
\author{Liang~Liu
        and~Ping~Wei%
\thanks{The authors are with the Center for Cyber Security, School of Electronic Engineering, University of
Electronic Science and Technology of China, Chengdu, 611731 China (e-mail:liu\_yinliang@outlook.com; pwei@uestc.edu.cn).}}

\markboth{Journal of \LaTeX\ Class Files,~Vol.~11, No.~4, December~2012}%
{Shell \MakeLowercase{\textit{et al.}}: Bare Demo of IEEEtran.cls for Journals}

\maketitle

\begin{abstract}
In this letter, we apply previous array receiver architecture which employs time-domain sub-Nyquist sampling techniques to jointly estimate frequency and direction-of-arrival(DOA) of narrowband far-field signals. Herein, a more general situation is taken into consideration, where there may be more than one signal in a subband. We build time-space union model, analyze the identification of the model, and give the maximum signal number which can be classified. We also proof that the Cram\'{e}r\text{-}Rao Bound (CRB) is lower than that of which employs Nyquist sampling. Simulation results verify the capacity to estimate the number of sources. Meanwhile, simulations show that our estimation performance closely matches the CRB and is superior for more sources than sensors, especially when the minimum redundancy array (MRA) is employed.
\end{abstract}


\IEEEpeerreviewmaketitle

\section{Introduction}
%
%\IEEEPARstart{J}{oint} estimation of carrier frequency and direction of arrival (DOA) for multiple signals has been found in many practical applications. For example, Cognitive Radio (CR) technique might be a good way to cope with the problem of the spectral congestion~\cite{Haykin2005, Yucek2009, Mishali2011a, Sun2013, Cohen2014}. One of the most important functions of CRs is to detect locally idle spectrum and then reactive the spectrum access from the concept of spectrum sensing. Generally, there exist three main spectrum, i.e., time, frequency and space. With the introduction and development of array processing techniques~\cite{Krim1996, Schmidt1986, Roy1986}, the spatial spectrum or DOA of a signal can be thought of as a new opportunity to improve the performance of CRs. Therefore, researchers have pay more attention to how to jointly estimate carrier frequencies and their DOAs of multiple signals~\cite{Lemma1998, Lemma2003}. Unfortunately, both of them exist at least two shortcomings. One is the pair matching problem for the carrier frequencies associated with the DOAs. The other is that time domain sampling rate is equal to or large than Nyquist sampling rate which will be considered as a bottleneck for wideband signal processing by CRs. For instance, It leads to prohibitive Nyquist sampling rate and its corresponding sampling data to be processed if the spectrum needed to be monitoring is from 300 MHz to several GHz~\cite{Haykin2005, Yucek2009, Mishali2011a, Sun2013, Cohen2014}.
\IEEEPARstart{T}{o} better deal with the problem of the spectral congestion~\cite{Haykin2005, Yucek2009, Mishali2011a, Sun2013, Cohen2014}, Cognitive Radio (CR) technique consider three main spectrum, i.e., time, frequency and space, to improve access possibility. The development of array processing techniques~\cite{Krim1996, Schmidt1986, Roy1986} provides a spatial spectrum access opportunity to increase the spectral resource utilization. Since space-time spectrum sensing need to make full use of spectral resource, jointly estimate carrier frequency and  direction-of-arrival(DOA)~\cite{Lemma1998, Lemma2003} is main challenge. There are two dramatical  shortcomings of the existing methods. One is the pair matching problem between the carrier frequencies and the DOAs. The other is that the Nyquist sampling is too high for wideband signal processing, which becomes a bottleneck for CRs.
Typically, it is intolerable due to of the too high Nyquist sampling rate or too much data to process when the monitoring range is from 300 MHz to several GHz~\cite{Haykin2005, Yucek2009, Mishali2011a, Sun2013, Cohen2014}.

Recent years, sub-Nyquist sampling technique has been widely researched to reconstruct a multiband signal from the data obtained under the Nyquist sampling rate~\cite{Mishali2011,Mishali2010,Eldar2009,Mishali2009}. Inspired by the idea, some joint estimation methods of DOA and carrier frequency were proposed for the under Nyquist sampling rate situation. The authors of~\cite{Ariananda2013} suggested a new structure, where each output of a linear array employs the multi-coset sampling. If the array is a nonuniform linear array, the signal can be compressed in both the time domain and spatial domain. \cite{Kumar2014} use only one additional identical delayed channel for each antenna to simplify the hardware complexity. Herein, the mismatch problem will happen when using an underlying uniform linear array (ULA). To this end, the authors proposed another structure in~\cite{ Kumar2015}, which still has low hardware complexity. In ~\cite{Kumar2016}, the authors proposed the so-called space-time array to jointly estimate frequency and DOA when the number of sources is more than the number of sensors. More recently, two joint DOA and carrier frequency recovery approaches based on the L-shaped ULAs were presented in~\cite{Stein2015}. In \cite{Liu2016}, a  receiver architecture and estimation algorithms is proposed to jointly estimate frequency and DOA. However, merely a special situation is considered in \cite{Liu2016}, where there is at most one signal in a subband.

Based on the  receiver architecture proposed in \cite{Liu2016}, this letter considers a more general situation, where there may exist more than one narrowband far-field signal in a subband. We built the time-space union model for this scenario.  After analyzing the identification of the model,  the maximum signal number which can be classified  is given. The proposed model is capable of estimating more signals than sensors. We also proof that the Cram\'{e}r\text{-}Rao Bound (CRB) is lower than that which employing Nyquist sampling. The simulation verifies the capacity to estimate the number of sources. Besides, simulation shows that our estimation performance closely matches the CRB and is superior for more sources than sensors, especially when the minimum redundancy array (MRA) is employed.
%This paper is organized as follows: in Section II, we describe the basic array signal model and point out the objective of this paper. In Section III, the proposed receiver architecture is presented, and a new  signal reception model is derived. In Section IV, two joint DOA and frequency estimation methods for the receiver architecture are proposed. In Section V, we deduce the corresponding CRB and demonstrate some result on CRB. Section VI carries out the simulation experiment and finally the conclusions of this paper are given in Section VII.

\section{Signal model and objective}
%In this section, we will give the array signal model and fundamental assumptions as well as the objective of this paper is also described.

%\subsection{Array signal model}

%\begin{align}\label{eqnArray}
%{\bf{x}}\left( t \right) = {\bf{As}}\left( t \right) + {\bf{n}}\left( t \right),
%\end{align}
%where ${\bf{x}}\left( t \right) = {\left[ {{x_1}\left( t \right), \cdots ,{x_M}\left( t \right)} \right]^{\rm{T}}}$ is the measurement vector, ${\bf{s}}\left( t \right) = {\left[ {{s_1}\left( t \right), \cdots ,{s_K}\left( t \right)} \right]^{\rm{T}}}$ is the vector of all signal values, ${\bf{n}}\left( t \right)= {\left[ {{x_1}\left( t \right), \cdots ,{x_M}\left( t \right)} \right]^{\rm{T}}} $ is the noise vector, which subjects to the zero-mean circular complex Gaussian distribution with covariance matrix  ${\sigma}^2 {\bf{I}}_M$. ${{{A}}_{mk}} = \exp \left( { - j{\phi _k}\left( {m - 1} \right)} \right)$ is the $(m,k)$th element of the steer array, where spatial phase
%\begin{align}\label{Phi}
%{\phi _k} = \frac{{2\pi d\sin \left( {\theta _k} \right)}}{c /{{f_k}}},
%\end{align}
%where ${\theta _k}$ and ${f_k}$ are the DOA and the center frequency of ${s_k}\left( t \right)$, respectively.

Consider $K$ narrowband far-field signals ${\bf{s}}\left( t \right) = {\left[ {{s_1}\left( t \right), \cdots ,{s_K}\left( t \right)} \right]^{\rm{T}}}$ impinging on a ULA composed of $M$ sensors, where ${\left(  \cdot  \right)^{\rm T}}$ denotes the transpose. It should be noted that arbitrary array form can be employed as explained in \cite{Liu2016}. The frequency domain array output can be written as \cite{Krim1996}
\begin{align}\label{eqnDOAfre}
{\bf{X}}\left( f \right) = {\bf{AS}}\left( f \right) + {\bf{N}}\left( f \right),
\end{align}
where ${{{A}}_{mk}} = \exp \left( { - j{\phi _k}\left( {m - 1} \right)} \right)$ is the $(m,k)$th element of the steer array. The spatial phase is shown as
\begin{align}\label{Phi}
{\phi _k} = \frac{{\pi {d}\sin \left( {{\theta _k}} \right){f_k}}}{{{f_N}}},
\end{align}
where $d$ is the distance between two consecutive antennas in half-wavelengths corresponding to the Nyquist sampling rate ${f_N}$. ${\theta _k}$ and ${f_k}$ are the DOA and the center frequency of ${s_k}\left( t \right)$, respectively. The sensor position vector is ${\bf{d}} = \left[ {0,1, \cdots ,M - 1} \right]d$. ${\bf{X}}\left( f \right) = {\left[ {{X_1}\left( f \right), \cdots ,{X_M}\left( f \right)} \right]^{\rm{T}}}$, ${\bf{S}}\left( f \right) = {\left[ {{S_1}\left( f \right), \cdots ,{S_K}\left( f \right)} \right]^{\rm{T}}}$, and ${\bf{N}}\left( f \right) = {\left[ {{N_1}\left( f \right), \cdots ,{N_M}\left( f \right)} \right]^{\rm{T}}}$ are the frequency domain expression of ${\bf{x}}\left( t \right)$, ${\bf{s}}\left( t \right)$, ${\bf{n}}\left( t \right)$, respectively.  ${X_m}\left( f \right)$ is the Fourier transform of ${x_m}\left( t \right)$.
The measurement vector and  the noise vector are defined as ${\bf{x}}\left( t \right) = {\left[ {{x_1}\left( t \right), \cdots ,{x_M}\left( t \right)} \right]^{\rm{T}}}$ and ${\bf{n}}\left( t \right)= {\left[ {{n_1}\left( t \right), \cdots ,{n_M}\left( t \right)} \right]^{\rm{T}}} $, respectively. The noise subjects to the zero-mean complex spatially and temporarily white Gaussian distribution with covariance matrix  ${\sigma}^2 {\bf{I}}_M$, where ${{\bf{I}}_M}$ stands for an $M \times M$ identity matrix.

The objective of this  letter is to simultaneously estimate the carrier frequency ${f_k}$ and DOA ${\theta _k}$ of multiple signals ${s_k(t)}$ under the Nyquist sampling rate.

\section{Signal reception model and Identification} \label{SectionModel}
%Now, we modify the traditional array signal receiver architecture and introduce the sub-Nyquist sampling technique into the architecture to reduce sampling rate. In this section, a novel architecture is presented and the corresponding signal reception model is derived.
%\subsection{Receiver architecture}
%Our receiver architecture is shown in Fig.\ref{figArcFull}. We apply multi-coset sampling \cite{Mishali2009} in Fig.\ref{figArcFull} as representative of sub-Nyquist sampling technology.
%In Fig.1., there are $M$ sensors and every sensor is followed by $P$ delay branches. All the ADCs are well-synchronized and sample at a sub-Nyquist sampling rate of ${f_{sub}} = {{{f_N}} / L}$, where ${f_N} = {1 / {{T_N}}}$ is the Nyquist sampling rate. The constant set $C=[c_1,c_2,\cdots,c_P]$ is referred to the sampling pattern where $0 \le {c_1} < {c_2} <  \cdots  < {c_P} \le L - 1$. ${y_{mp}}\left[ n \right]$ denotes the sampled signal corresponding to the $m$th sensor, $p$th branch.
%The  average sampling rate of the multi-coset sampling is
% \begin{align}\label{feq}
%{f_E} = \frac{{P{f_N}}}{L},
%\end{align}
%which is lower than the Nyquist rate $f_N$ when $P<L$.

%\begin{figure}[!t]
%\centering
%\includegraphics[width=2.5in]{A1.eps}
%\caption{Proposed receiver architecture.}
%\label{figArcFull}
%\end{figure}

\subsection{Signal reception model}
We employ the  receiver architecture as Fig. 1 in \cite{Liu2016}. There are $M$ sensors and every sensor is followed by $P$ delay branches in the architecture. ${f_{sub}} = {{{f_N}} / L}$ is the sub-Nyquist sampling rate, where $L$ is the sampling rate reduction factor and ${y_{mp}}\left[ n \right]$ denotes the sampled signal corresponding to the $m$th sensor, $p$th branch. The sampling pattern is set at $C=[c_1,c_2,\cdots,c_P]$. In (\ref{eqnDOAfre}), suppose that column order of $\bf{A}$ is determined by the frequency of signals, and there are ${K_l}$ signals in $l$th subband. The following  notations are introduced: ${{\bf{S}}^{\rm T}}\left( f \right){\rm{ = }}\left[ {{\bf{S}}_1^{\rm T}\left( f \right), \cdots ,{\bf{S}}_L^{\rm T}\left( f \right)} \right] \in {C^{1 \times K}}$,  ${\bf{S}}_l^{\rm T}\left( f \right) = \left[ {{S_{l1}}\left( f \right), \cdots ,{S_{l{K_l}}}\left( f \right)} \right] \in {C^{1 \times {K_l}}}$, $f \in \left[ {0,\frac{1}{T}} \right)$, $K = \sum\nolimits_{l = 1}^L {{K_l}} $, where ${{\bf{S}}_l}\left( f \right)$ refer to  all the signals in the $l$th subband, ${S_{lk}}\left( f \right)$ is the $k$th signal in the $l$th subband. The steel vector will be ${\bf{A}}{\rm{ = }}\left[ {{{\bf{A}}_1}, \cdots ,{{\bf{A}}_L}} \right] \in {C^{M \times K}}$, $ {{{\bf{A}}_l}}  = \left[ {{{\left( {{{\bf{A}}_l}} \right)}_1}, \cdots ,{{\left( {{{\bf{A}}_l}} \right)}_{{K_l}}}} \right] \in {C^{M \times K_l}} $, where ${{\bf{A}}_l}$ is the steel matrix corresponding  to ${{\bf{S}}_l}\left( f \right)$ as well as  the $l$th block of ${\bf{A}}$, and ${{\left( {{{\bf{A}}_l}} \right)}_k}$ is the steel vector corresponding  to ${S_{lk}}\left( f \right)$  as well as the $k$th column of $\left( {{{\bf{A}}_l}} \right)$.

According to (\ref{eqnDOAfre}), the output of the $m$th sensor is
\begin{align}\label{eqnunionEl}
{X_m}\left( f \right) = {{\bf{A}}^m}{\bf{S}}\left( f \right) + {N_m}\left( f \right),f \in \left[ {0,\frac{1}{T_N}} \right),
\end{align}
 where ${{\bf{A}}^m}{\rm{ = }}\left[ {{{\left( {{{\bf{A}}_1}} \right)}^m}, \cdots ,{{\left( {{{\bf{A}}_L}} \right)}^m}} \right]$ is the $m$th row of ${{\bf{A}}}$.
Combining the conclusions of \cite{Mishali2009, Liu2016}, the output of all branches of $m$th sensor is expressed as
\begin{align}\label{eqnSubNy}
{{\bf{Y}}_m}\left( f \right) = {\bf{B}}{\overline {\bf{X}} _m}\left( f \right), f \in \mathcal{F}
\end{align}
where ${{{B}}_{il}} = \frac{1}{{\sqrt L }}\exp \left( {j\frac{{2\pi }}{L}{c_i}l} \right)$,
${{\bf{Y}}_m}\left( f \right) = \sqrt L T_N\left[ {{Y_{m1}}\left( {{e^{j2\pi fT_N}}} \right), \cdots ,{Y_{mP}}\left( {{e^{j2\pi fT_N}}} \right)} \right]^{\rm{T}}$,
 ${\overline {\bf{X}} _m}\left( f \right) = {\left[ {{X_{m1}}\left( f \right), \cdots ,{X_{mL}}\left( f \right)} \right]^{\rm{T}}}$,
 ${X_{ml}}\left( f \right) = {X_m}\left( {f + \left(l-1\right){f_{sub}}} \right)$,
$\mathcal{F} \buildrel \Delta \over = \left[ {0,{f_{sub}}} \right)$.
${T_N}{\rm{ = }}{1 \mathord{\left/
 {\vphantom {1 {{f_N}}}} \right.
 \kern-\nulldelimiterspace} {{f_N}}}$ is the Nyquist sampling interval.
${Y_{mp}}\left( {{e^{j2\pi fT_N}}} \right)$ is  the discrete-time Fourier transform of ${y_{mp}}\left[ n \right]$.
Based on (\ref{eqnunionEl}) and the form of ${\overline {\bf{X}} _m}\left( f \right)$, we have
\begin{align}\label{eqnArrayFreSub}
{\overline {\bf{X}} _m}\left( f \right)= {}&{\mathop{\rm blkdiag}\nolimits} \left( {{{\left( {{{\bf{A}}_1}} \right)}^m}, \cdots ,{{\left( {{{\bf{A}}_L}} \right)}^m}} \right)\overline {\bf{S}} \left( f \right)+ {\overline {\bf{N}} _m}\left( f \right), \nonumber\\
&{}f \in {\cal F},
\end{align}
where ${\overline {\bf{S}} ^{\rm T}}\left( f \right)= \left[ {\overline {\bf{S}} _1^{\rm T}\left( f \right), \cdots ,\overline {\bf{S}} _L^{\rm T}\left( f \right)} \right]$, $\overline {\bf{S}} _l^{\rm T}\left( f \right) = \left[ {{{\overline S }_{l1}}\left( f \right), \cdots ,{{\overline S }_{l{K_l}}}\left( f \right)} \right]$, ${\overline S _{lk}}\left( f \right){\rm{ = }}{S_{lk}}\left( {f + \left( {l - 1} \right){f_{sub}}} \right)$, ${\overline {\bf{N}} _m}\left( f \right) = {\left[ {{{\overline N }_{m1}}\left( f \right), \cdots ,{{\overline N }_{mL}}\left( f \right)} \right]^{\rm T}}$, ${\overline N _{ml}}\left( f \right) = {N_m}\left( {f + \left( {l - 1} \right){f_{sub}}} \right)$. ${\mathop{\rm blkdiag}\nolimits} \left( {{{\bf{z}}_1}, \cdots ,{{\bf{z}}_m}} \right)$ represents a block diagonal matrix with diagonal entries ${{{\bf{z}}_1}, \cdots ,{{\bf{z}}_m}}$.

\noindent Substituting (\ref{eqnArrayFreSub}) into (\ref{eqnSubNy}), we get
\begin{align}\label{eqnOneCh}
{{\bf{Y}}_m}\left( f \right)= &  {\bf{B}}{\mathop{\rm blkdiag}\nolimits} \left( {{{\left( {{{\bf{A}}_1}} \right)}^m}, \cdots ,{{\left( {{{\bf{A}}_L}} \right)}^m}} \right)\overline {\bf{S}} \left( f \right) + {\bf{B}}{\overline {\bf{N}} _m}\left( f \right) \nonumber  \\
 = &  \left[ {{{\left( {{{\bf{A}}_1}} \right)}^m} \otimes {{\bf{B}}_1}, \cdots ,{{\left( {{{\bf{A}}_L}} \right)}^m} \otimes {{\bf{B}}_L}} \right]\overline {\bf{S}} \left( f \right)  \nonumber \\
&+ {\bf{B}}{\overline {\bf{N}} _m}\left( f \right), f \in \mathcal{F}
\end{align}
where $\overline {\bf{S}} \left( f \right) = {\left[ {\overline {\bf{S}} _1^{\rm{T}}\left( f \right), \cdots ,\overline {\bf{S}} _K^{\rm{T}}\left( f \right)} \right]^{\rm{T}}}$, ${\overline {\bf{S}} _k}\left( f \right) = {\left[ {{S_{k1}}\left( f \right), \cdots ,{S_{kL}}\left( f \right)} \right]^{\rm{T}}}$. $ \otimes$ denotes the Kronecker product.
%{{{\bf{Y}}_m}\left( f \right) &= {\bf{B}}{\rm{blkdiag}}\left( {{{\left( {{{\bf{A}}_1}} \right)}_m} \cdots {{\left( {{{\bf{A}}_L}} \right)}_m}} \right)\overline {\bf{S}} (f) + {{\bf{N}}_m}\left( f \right)}\\
% &= \left[ {{{\left( {{{\bf{A}}_1}} \right)}_m} \otimes {{\bf{B}}_1} \cdots {{\left( {{{\bf{A}}_L}} \right)}_m} \otimes {{\bf{B}}_L}} \right]\overline {\bf{S}} (f) + {{\bf{N}}_m}\left( f \right)
% \\
% &{1 \le m \le M,f \in F}.

\noindent Then, combining all $m$ can result in
\begin{align}\label{Yf}
{\bf{Y}}\left( f \right) &= \left[ {{{\bf{A}}_1} \otimes {{\bf{B}}_1}, \cdots ,{{\bf{A}}_L} \otimes {{\bf{B}}_L}} \right]\overline {\bf{S}} (f) + \left( {{{\bf{I}}_M} \otimes {\bf{B}}} \right)\overline {\bf{N}} \left( f \right), \nonumber \\% \label{Yf2}
& = {\bf{G}}\overline {\bf{S}} (f) + {{\bf{I}}_{\bf{B}}}\overline {\bf{N}} \left( f \right),f \in {\cal F}
\end{align}
where ${\bf{Y}}\left( f \right) = {\left[ {{\bf{Y}}_1^{\rm{T}}\left( f \right), \cdots ,{\bf{Y}}_M^{\rm{T}}\left( f \right)} \right]^{\rm{T}}}$, $\overline {\bf{N}} \left( f \right){\rm{ = }}\left[ {{{\overline {\bf{N}} }_1}\left( f \right), \cdots ,{{\overline {\bf{N}} }_L}\left( f \right)} \right]$. Actually, ${\bf{Y}}\left( f \right)$ in (\ref{Yf}) is the  matrix form of  the output of all branches of all sensors.

%\begin{remark}
%It is clear that (\ref{eqnSubNy}) and (\ref{eqnunionEl}) are the sub-Nyquist sampling model and DOA model, respectively. Note that the information  in frequency domain can be obtained from (\ref{eqnSubNy}). Equation (\ref{Yf}) is the unified model, where the frequency domain and the spatial domain are treated equally. Then we can obtain the information of frequency domain and spatial domain from this unified model simultaneously rather than separately or sequentially. In \cite{Ariananda2013, Kumar2014, Kumar2015}, the sub-Nyquist sampling is also applied to array receiver, but only separate  models are given. They do not give one unified model which simultaneously takes all branches' data and information in the frequency domain and the spatial domain into consider.
%\end{remark}

%In (\ref{Yf}), since the Kronecker product is used to unify the frequency domain and spatial domain into a two-dimensional matrix form, this equation can be viewed from the perspective of third-order tensor. We will discuss the third-order tensor in the next section. On the other hand, if we regarded ${\bf{G}}$ as  a special array manifold, subspace decomposition theory can be employed. Of course, different perspectives will derive different methods, which will be analyzed in detail.

After modeling the  reception model, we can apply JDFSD in \cite{Liu2016} to solve it. The difference between the two papers is that there  maybe be one or more signals in one subband in this paper while there is at most one signal in one subband in \cite{Liu2016}. Therefore, the method in this paper is named JDFSD4MU.

\subsection{Identification}

For identification, only consider a simple situation: the branch number is equal to the sampling rate reduction factor ($P=L$), and there are no more than $M-1$ signals in each subband, which are from different DOAs.  So we have $Rank({{\bf{V}}_i})=Rank({{\bf{A}}_i})={{K}_i}$, where ${\bf{V}}_i = {{\bf{A}}_i} \otimes {{\bf{B}}_i}$. $Rank(\cdot)$ denotes the rank of a matrix. Apparently, we will hold ${\bf{V}}_i \bot {\bf{V}}_j, i \neq j$, since ${{\bf{B}}^{\rm H}}{\bf{B}}{\rm{ = }}{\bf{I}}$ when $P=L$.  Further, we get $ Rank({\bf{G}}) = \sum_{i=1}^{L} Rank({{\bf{V}}_i}) = K$.
Based on the subspace decomposition theory \cite{Schmidt1986} and $Rank({\bf{G}}) = K$, the model (\ref{Yf}) can be solved and the maximum signal number  which can be classified is $\left( {M - 1} \right)L$.

\section{ Cram\'{e}r\text{-}Rao Bound} \label{SectionCRB}
%\subsection{Cram\'{e}r\text{-}Rao Bound}
Based on \cite{Liu2016}, the CRB of our model is given by
\begin{align}\label{CRBsubP}
{{\rm{CRB}}_{sub}} = \frac{{{\sigma ^2}}}{{2T}}{\left( {\Re \left( {\left( {{{\bf{E}}^{\rm{H}}}{{\bf{P}}_{\bf{G}}}{\bf{E}}} \right) \odot {\bf{R}}_{\bf{S}}^{\rm{H}}} \right)} \right)^{ - 1}}
\end{align}
where ${{\bf{P}}_{\bf{G}}} = {\bf{I}} - {\bf{G}}{{\bf{G}}^\dag }$ , ${\bf{E}} = \left[ {{{\bf{E}}_1}, \cdots ,{{\bf{E}}_K}} \right]$, ${{\bf{E}}_k} = \frac{{d{{{\bf{G}}_k}}}}{{d{\phi _k}}}$. ${\left(  \cdot  \right)^{\rm H}}$ and ${\left(  \cdot  \right)^{^\dag }}$  denote Hermitian transpose, and Moore-Penrose pseudo-inverse, respectively. ${\bf{R}}_{\bf{S}}=\rm blkdiag ({\bf{R}}_{{\bf{S}}_1},\ldots,{\bf{R}}_{{\bf{S}}_L})$ is signal autocorrelation matrix and {$T$} is the snapshots of observation.
%
%For comparing, the CRB  which employs the Nyquist sampling (marked as $\textrm{CRB}_{Ny}$) \cite{Stoica1990} is rewritten here.
%\begin{align}\label{CRBNyP}
%{\rm{CRB }}_{Ny}= \frac{{{\sigma ^2}}}{{2T}}{\left( {\Re \left( {\left( {{{\bf{D}}^{\rm{H}}}{{\bf{P}}_{{{\bf{A}}}}}{\bf{D}}} \right) \odot {\bf{R}_{\bf{S}}^{\rm{H}}}} \right)} \right)^{ - 1}}
%\end{align}
%where ${{\bf{P}}_{\bf{A}}} = {\bf{I}} - {\bf{A}}{{\bf{A}}^\dag }$, ${\bf{D}} = \left[ {{{\bf{D}}_1}, \cdots ,{{\bf{D}}_K}} \right]$, ${{\bf{D}}_i} = \frac{{d{{\bf{A}}_i}}}{{d{\phi _i}}}$.

Next, we will show that ${\rm{CRB}}_{sub}$ is lower than ${\rm{CRB}}_{Ny}$  under the same conditions: the same array arrangement,
same noise environment, and same snapshot($P = L$).
Considering the expression of ${\bf{G}}$ and ${\bf{E}}$ and ${{\bf{B}}^{\rm H}}{\bf{B}}{\rm{ = }}{\bf{I}}$ when $P=L$, we hold
 \begin{align}\label{EqBlk}
{{\bf{G}}^{\rm H}}{\bf{G}}&={\mathop{\rm blkdiag}\nolimits} \left( {{\bf{A}}_1^{\rm H}{{\bf{A}}_1}, \cdots, {\bf{A}}_L^{\rm H}{{\bf{A}}_L}} \right), \nonumber \\
{{\bf{E}}^{\rm H}}{\bf{E}}&={\mathop{\rm blkdiag}\nolimits} \left( {{\bf{D}}_1^{\rm H}{{\bf{D}}_1}, \cdots ,{\bf{D}}_L^{\rm H}{{\bf{D}}_L}} \right), \nonumber \\
{{\bf{E}}^{\rm H}}{\bf{G}}&={\mathop{\rm blkdiag}\nolimits} \left( {{\bf{D}}_1^{\rm H}{{\bf{A}}_1}, \cdots ,{\bf{D}}_L^{\rm H}{{\bf{A}}_L}} \right),
\end{align}
where ${\bf{D}} = \left[ {({{\bf{D}}_1}), \cdots ,({{\bf{D}}_L})} \right]$, ${{\bf{D}}_l} = \left[ { ({{\bf{D}}_l})_1, \cdots ,({{\bf{D}}_l})_{K_l}} \right]$, ${({\bf{D}}_l)_k} = \frac{{d{({\bf{A}}_l)_k}}}{{d{\phi _k}}}$.

Further, based on (\ref{EqBlk}) and after some matrix manipulations, we have
 \begin{align}\label{EqCRB1}
{{\rm{CRB}}_{sub}} = {\mathop{\rm blkdiag}\nolimits} \left( {{{\bf{C}}_1}, \cdots ,{{\bf{C}}_L}} \right),
\end{align}
where ${{\bf{C}}_l}=\frac{{{\sigma ^2}}}{{2T}}\Re {\left( {\left( {{\bf{D}}_l^{\rm H}{{\bf{P}}_{{{\bf{A}}_l}}}{{\bf{D}}_l}} \right) \odot {\bf{R}}_{{{\bf{S}}_l}}^{\rm{H}}} \right)^{ - 1}}$.
(\ref{EqCRB1}) shows that the new steer vectors corresponding to different subband would be  completely uncorrelated in spite of that the primary steer vectors are correlated. This is the reason why once the sub-Nyquist sampling is employed, the performance of DOA estimation for any one subband is not affected by other subbands.

We consider the performance of DOA estimation in the following situations:

I) All of the signals are distributed in the $l$th subband and there is no signal in the other subbands. If the Nyquist sampling is employed, the CRB for the DOA in the $l$th subband is
${\overline {\rm{CRB}} _{Ny}}\left( l \right) = \frac{{{\sigma ^2}}}{{2T}}\Re {\left( {\left( {{\bf{D}}_l^{\rm H}{{\bf{P}}_{{{\bf{A}}_l}}}{{\bf{D}}_l}} \right) \odot {\bf{R}}_{{{\bf{S}}_l}}^{\rm{H}}} \right)^{ - 1}}$.

II) The signals are distributed in not only the $l$th subband but also the other subbands. If the Nyquist sampling is employed, the CRB for the DOA in the $l$th subband is ${\rm{CRB}}_{Ny}\left( l \right)$.

III) The distribution of the signal is the same as II), but the sub-Nyquist sampling is employed. The CRB for the DOA in the $l$th subband is ${\rm{CRB}}_{sub}\left( l \right) = {{\bf{C}}_l} = \frac{{{\sigma ^2}}}{{2T}}\Re {\left( {\left( {{\bf{D}}_l^{\rm H}{{\bf{P}}_{{{\bf{A}}_l}}}{{\bf{D}}_l}} \right) \odot {\bf{R}}_{{{\bf{S}}_l}}^{\rm{H}}} \right)^{ - 1}}$.

Based on section V of \cite{Liu2016}, the increase of the number of DOA will degrade the performance of DOA estimate, so we hold ${\overline {\rm{CRB}} _{Ny}}\left( l \right) \preceq {\rm{CRB}}_{Ny}\left( l \right)$. Further, we have
 \begin{align}\label{EqCRBLow}
{\rm{CRB}}_{sub}\left( l \right) = {\overline {\rm{CRB}} _{Ny}}\left( l \right) \preceq {\rm{CRB}}_{Ny}\left( l \right).
\end{align}
(\ref{EqCRBLow}) shows that when the signals are distributed in not only the $l$th subband but also the other subbands, in terms of the performance of DOA estimate for the signals in the $l$th subband, the method which employs sub-Nyquist sampling is better than the method which employs Nyquist sampling. Because $l$ is arbitrary,  $ {\rm{CRB}}_{sub}\left( l \right)\preceq{\rm{CRB}}_{Ny}\left( l \right)$ is true for $1 \le l \le L$. So, we get
 \begin{align}\label{EqCRBCMP}
{\rm{CRB}}_{sub} \preceq {\rm{CRB}}_{Ny},
\end{align}
the equality holds if and only if all of the signals are distributed in same one subband.

\section{Simulation}

In this section, the numerical simulations are carried out to study the performance with different source number. In the simulations, some complex-valued narrowband far-field non-coherent signals with equal power imping on a ULA composed of $M = 7$ sensors which are separated by a half wavelength corresponding to Nyquist sampling rate, which would probably be the signal highest frequency. We employ MRA to compress the signal in the spatial domain. The MRA is composed of $M=7$ sensors which are located at ${\bf{d}} = \left[ {0,1,4,10,16,22,28} \right]d$. We fix the number of snapshots at $T = 7000$ for Nyquist sampling, $T_{sub} = T/L $ for sub-Nyquist sampling, the Nyquist sampling rate at $f_N = 10 $ GHz, the sampling rate reduction factor at $L = 7$, and the branch number at $P=L$.
We set $\bm{\vartheta} {\rm{ = }}\left[ {{\vartheta _1},{\vartheta _2}, \cdots ,{\vartheta _{14}}} \right]$, where ${\vartheta _i}$ follows the uniform distribution between ${{-60}^\circ }$ and ${{60}^\circ }$, where $1 \le i \le 14$.
We set $\bm{\upsilon}=\left[ {{\upsilon_1},{\upsilon_2}, \cdots ,{\upsilon_{14}}} \right]$, where ${\upsilon_i},{\upsilon_{i + 7}}$ are in the ${q_i}$-th subband, where $1 \le i \le L$, and $\left\{ {{q_i}} \right\} = \left\{ {1,2, \cdots ,L} \right\}$.
We set DOA ${\bm{\theta }} = \left[ {{\vartheta _1},{\vartheta _2},\cdot\cdot\cdot,{\vartheta _K}} \right]$ and frequency ${\bf{f}} = \left[ {{\upsilon_1},{\upsilon_2},\cdots,{\upsilon_K}} \right]$. The definition of signal-to-noise ratio (SNR) and  root-mean-square error (RMSE) of DOA is same as \cite{Liu2016}. The \textrm{SNR} is fixed at 20 dB. 2000 Monte Carlo trials are implemented.

\begin{figure}[!t]
\centering
\includegraphics[width=3.0in]{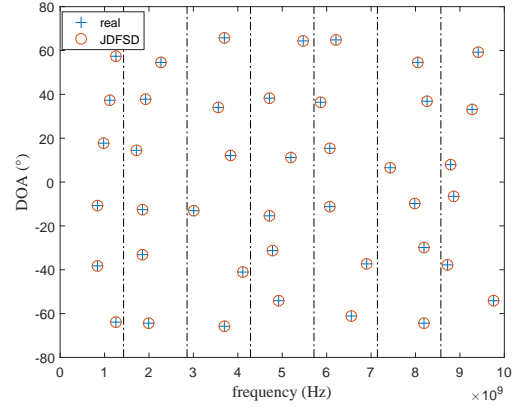}
\caption{Actual and estimated frequencies and their DOAs,  maximum signal number $K=42$ when $M=7,P=L=7$.}
\label{figMaxK}
\end{figure}

The first simulation will verify the estimation capacity. Based on section \ref{SectionModel}, the maximum signal number which can be classified is $K=(M-1)L=42$. Fig. \ref{figMaxK} shows that the frequencies and DOAs can be accurately estimated when noise is free. The identical targets are at most $M-1 = 6$ in each subband with $M=7$ sensors.

\begin{figure}[!t]
\centering
\includegraphics[width=3.0in]{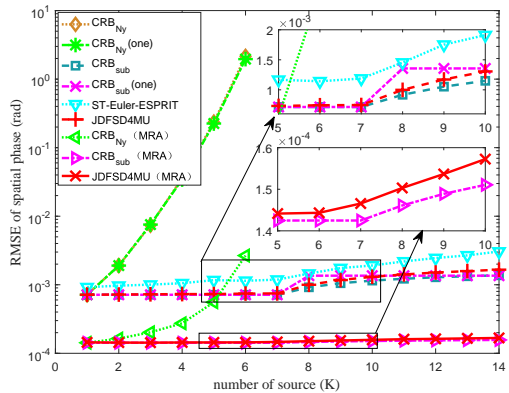}
\caption{RMSE of phase estimates versus number of source.}
\label{figPK}
\end{figure}

As the analysis in part A of section VI in \cite{Liu2016}, the joint estimation performance is limited by the spatial phase or DOA estimation performance. Hence, only the spatial phase estimation performance is given in the simulation. Meanwhile, we will compare our methods with ST-Euler-ESPRIT in~\cite{Kumar2016}. The receiver configuration parameters of ST-Euler-ESPRIT are the same as ours. The delay is Nyquist sampling interval $T_N=1/f_N$.
Fig.\ref{figPK} shows that the DOA estimation performances of algorithm JDFSD4MU is close to $\textrm{CRB}_{sub}$ and lower than $\textrm{CRB}_{Ny}$ except $K=1$ whether ULA or MRA is employed. Apparently, when the MRA is employed, the spatial estimation performance is improved.
When $K\leq L$, $\textrm{CRB}_{sub}$  and JDFSD4MU are not influenced by the signal number.
When $K \geq L$, the traditional structure can not obtain the estimation of DOAs.
However, our method still can achieve the estimation although  the $\textrm{CRB}_{sub}$  increases  with signal number.
The trend of $\textrm{CRB}_{sub}(one)$ shows that increasing signal number only influences the estimation performance of the targets  which are in the same subband, where $\textrm{CRB}_{sub}(one)$ is the CRB of the DOA in the $p_i$-th subband. However, it does not happen to $\textrm{CRB}_{Ny}(one)$. $\textrm{CRB}_{Ny}$  increases with the signal number, and increases faster than exponential function of the signal number. Those meet the analysis in section \ref{SectionCRB}. As for ST-Euler-ESPRIT, the RMSE increase with signal number and the performance is inferior to JDFSD4MU. ST-Euler-ESPRIT is limited by ULA, so the spatial estimation performance can not be improved by changing the array form as JDFSD4MU.

\section{Conclusions}
In this  letter, for the scenario where there may be more than one signal in a subband, by applying previous array receiver architecture and employing sub-Nyquist sampling techniques, we derived a more general time-space union model to jointly estimate frequency and DOA. We analyzed the identification of the model and gave the maximum signal number which can be classified so that the proposed model is capable to estimate more signals than sensors. We also proved that the CRB is lower than that employ Nyquist sampling. Furthermore, the simulation results verify the conclusions about the identification and CRB. Besides, the MRA can be employed to compress the signal in the spatial domain and improve the spatial estimation performance.

\ifCLASSOPTIONcaptionsoff
  \newpage
\fi

\bibliographystyle{IEEEtran}
\bibliography{IEEEabrv,Ref}

\end{document}